# Wavefront Sensor for Laser Beams Based on Reweighted Amplitude Flow Algorithm

Ondřej Denk[1,2,*], Jan Pilař[1], Martin Divoký[1], Miroslav Čech[2] and Tomáš Mocek[1]

1. Hilase Centre, Institute of Physics, ASCR, Za Radnici 828, 252 41, Dolni Brezany, Czech Republic
2. Faculty of Nuclear Sciences and Physical Engineering, Czech Technical University in Prague, Brehova 7, 115 19, Prague 1, Czech Republic
* *Ondrej.denk@hilase.cz

**Abstract**

We present a reference-free computational wavefront sensor based on binary amplitude modulation and phase retrieval. The method employs Digital Micro-mirror Device as a programmable amplitude modulator and reconstructs the complex optical field from multiple far-field intensity measurements using the Reweighted Amplitude Flow algorithm with Optimal Spectral Initialization. Unlike classical pupil-plane wavefront sensors, the proposed architecture does not include any wavelength-specific optical elements, enabling straightforward adaptation across a broad spectral range. The achievable spatial resolution of the reconstructed wavefront is scalable with the modulator resolution. We experimentally demonstrate wavefront reconstruction at 650 nm and at 2116 nm, where commercial wavefront sensors are not widely available. The reconstructed wavefront is validated against a commercial lateral shearing interferometer at 650 nm, and the method is further integrated into a closed-loop adaptive optics system using a deformable mirror. The approach is particularly suited for applications requiring high spatial resolution and large dynamic range in slowly varying or quasi-static laser fields, where computational reconstruction speed is not of the primary concern.

## 1. Introduction

Control of the laser beam quality is a crucial task in many optical laboratories and high-power laser facilities where complex optical chains and numerous transmissive and reflective components introduce cumulative aberrations [1–3]. These aberrations lead to sub-optimal beam propagation performance, give rise to intensity inhomogeneities, and act as a source for the formation of parasitic hot spots, potentially resulting in optical damages [4,5]. Control of the laser beam wavefront can prevent such unwanted phenomena and allows control over the beam intensity profile at the target.

An optical field is fully characterized by its amplitude and phase. Because of the high frequencies of visible and infrared light (~$10^{15}$ Hz), the phase information is typically lost during the image acquisition process due to signal integration. Conventional image sensors measure only time-averaged intensity, thereby discarding phase information. This limitation motivates the development of wavefront sensing techniques capable of reconstructing the phase of coherent optical fields from intensity-only measurements [6].

Multiple wavefront acquisition principles have been proposed [7] that allow measurement of the phase information along with the intensity. These techniques can be divided into two categories according to the basic principle of operation [8]. The pupil plane techniques, being the most prominent ones, include Shack-Hartmann sensor [9–11], Pyramid sensor [12,13] and Shearing Interferometry sensor [14,15]. These techniques, however, have limitations in dynamic range and/or spatial resolution which require special treatment to overcome [16,17]. The pupil plane images are typically obtained by employing designated refractive optical elements that affect the spatial phase profile, which can be subsequently extracted. In particular, the Shearing Interferometer systems achieve high wavefront resolution but can suffer from bulkiness caused by the complexity of the system, promoting increased sensitivity to vibrations on a submillimeter scale [18].

An alternative class of wavefront sensing techniques exploits redundancy in intensity measurements acquired in one or more planes perpendicular to the optical axis to computationally recover the phase. These methods include curvature sensing, phase diversity, and general phase retrieval approaches [19–25]. In contrast to classical pupil-plane sensors, computational methods typically offer a larger dynamic range and are not inherently limited by the spatial sampling of lenslet arrays or interferometric fringes. Spatial resolution of the reconstructed wavefront is given by the spatial resolution of the modulating or recording device. Primary limitation is computational complexity and sensitivity to noise, which often restrict their applicability to slowly varying or static fields.

The phase retrieving algorithms have a long history dating back to 1972 when R. W. Gerchberg and W. O. Saxton [26] presented their Error Reduction algorithm (ER) that uses phase-less intensity distribution of near- and far-field image, i.e. two Fourier coupled planes. The authors have also shown that the cumulative error in phase does not increase in the case of noiseless data but doesn't necessarily converge to an optimal solution.

A particularly active area within computational phase retrieval is the use of binary amplitude modulation, frequently implemented using Digital Micromirror Devices (DMDs), to encode phase information into multiple intensity measurements. Early demonstrations showed that random or pseudorandom binary masks enable recovery of the complex optical field from far-field intensity measurements [27]. Subsequent work explored ptychography-inspired formulations [28], systematic amplitude modulation strategies [29], and optimized binary mask design to improve noise robustness and reduce the number of required measurements [30,31]. These studies demonstrate that mask design plays a decisive role in reconstruction fidelity, stability, and speed. Within this context, the present work focuses on the system-level realization and experimental validation of a reference-free wavefront sensing architecture that combines binary amplitude modulation with the Reweighted Amplitude Flow algorithm, emphasizing wavelength versatility, high spatial resolution, and compatibility with closed-loop adaptive optics application.

In this paper we report on the use of Amplitude flow algorithm for Phase retrieval computations Utilizing Coded Aperture Masking (APUCAM). This method allows the development of a DMD-based amplitude-modulation wavefront sensor using Reweighted Amplitude Flow with Optimal Spectral Initialization (RAF-OSI) as the reconstruction engine. The focus of this work is the demonstration of high-resolution wavefront reconstruction with spatial sampling scalable only with the modulator resolution and extension of the method to the near infrared at 2116 nm, where high-resolution wavefront sensors are not widely available. A significant advantage of the method is its reliance on intensity-only detection with no refractive elements modulating the beam. Omitting wavelength specific phase-type light modulation allows the system to be readily adapted for a wide range of laser wavelengths.

We test the functionality of the method by comparing wavefronts reconstructed by APUCAM and by a commercially available lateral shearing interferometer wavefront sensor (SID4 by Phasics S.A.). The sensor is further integrated into a closed-loop adaptive optics system including a deformable mirror. The mirror is characterized using the sensor and successful operation of the closed loop is demonstrated. The proposed measurement method is robust and easily applicable as a practical, wavelength-versatile diagnostic and AO feedback tool for laboratory environments where high spatial resolution and large dynamic range are prioritized over real-time operation.

## 2. Principles of Computational Reconstruction

For clarity and reproducibility, we explicitly define the forward model used to generate intensity measurements. Let

$$u(x,y) = A(x,y)\exp[i\phi(x,y)] \tag{1}$$

denote the complex optical amplitude in the DMD plane. The field is modulated by an elementwise multiplication of a binary amplitude mask $M_i(x,y) \in \{0,1\}$, producing

$$u_i(x,y) = M_i(x,y) \odot u(x,y). \tag{2}$$

Using the paraxial and the scalar diffraction approximations, the complex field in the focal plane of a lens with focal length $f$ is given by the Fraunhofer diffraction integral, which is proportional to the Fourier transform of the field in the DMD plane:

$$U_i(\xi,\eta) = \mathcal{F}\{u_i(x,y)\}(\xi,\eta), \tag{3}$$

Under Fraunhofer propagation by a lens of focal length $f$, the focal-plane spatial coordinates $(X,Y)$ relate to spatial frequencies via $\xi = X/(\lambda f)$, $\eta = Y/(\lambda f)$, where $\lambda$ is the optical wavelength.

The recorded intensity measurement is then

$$y_i[p] = |U_i(X_p, Y_p)|^2 + n_i[p], \tag{4}$$

where $n_i$ represents measurement noise and $p$ indexes camera pixels within the area of interest.

The phase retrieval problem then takes the form of extracting amplitude $u(x,y)$ from a set of intensity-only measurements $\{y_i\}_{i=1}^m$ corresponding to different binary amplitude masks $\{M_i\}$.

Similar problems arise in e.g. X-ray crystallography [32,33], astronomy [34,35] and ptychography [36]. The problem of solving a system of quadratic equations can be defined as empirical loss minimization problem

$$\hat{u} = \arg\min_{u \in \mathbb{C}^N} \frac{1}{2m} \sum_{i=1}^m \||\mathcal{F}\{M_i \odot u\}| - \sqrt{y_i}\|_2^2, \tag{5}$$

where $\hat{u} \in \mathbb{C}^n$ is the optimal solution to the minimization problem.

This presents a non-convex problem that is non-deterministic polynomial-time hard (NP-hard) and thus inappropriate to computationally solve using e.g. least squares estimate.

The ER algorithm solves the minimization iteratively via Fourier propagation of the intensity distributions between the focal and pupil planes while conserving the reconstructed wavefront between the iteration steps. This procedure is repeated until a convergence criterion is met, and the complex field is obtained. Based on the ER algorithm principle, more phase retrieval algorithms appeared [37–41]. These algorithms are less prone to stagnating in a local minimum compared to ER.

Phase diversity and Curvature sensor algorithms use similar principles, although they use intensity measurements from multiple non-conjugated planes that are not directly related via Fourier transformation as denoted in Eq. 3. One example is Wirtinger

Intensity Flow algorithm [42]. Phase Diversity algorithms are frequently combined with gradient descent optimization of predefined aberration set in the Fourier domain [43].

Others [44–46] used the pseudorandom intensity modulation to first alter the near field intensity distribution, encode the phase and thus create an abundance of data. A relatively small number of recorded focal plane images of a beam with amplitude-modulated near field allows estimation of the wavefront shape with high fidelity. Related state-of-the-art computational methods developed for ptychography [47], also rely on the mathematical principle of extracting phase from multiple intensity measurements and have been applied in high-resolution laser wavefront sensing [48]. Our proposed approach, which employs Coded Aperture Masking with DMD, offers a simpler, compact and versatile alternative.

In this paper, we take up an approach of pseudorandom intensity modulation by a set of complementary binary masks. For the reconstruction, we utilize the Reweighted Amplitude Flow algorithm with Optimal Spectral Initialization [49] (RAF-OSI). It is suggested that a minimum of *2n - 1* measurements are required for the reconstruction of *n* points of noiseless data [46]. In the case of data containing noise, the necessary number of measurements is increased.

## 3. Methods

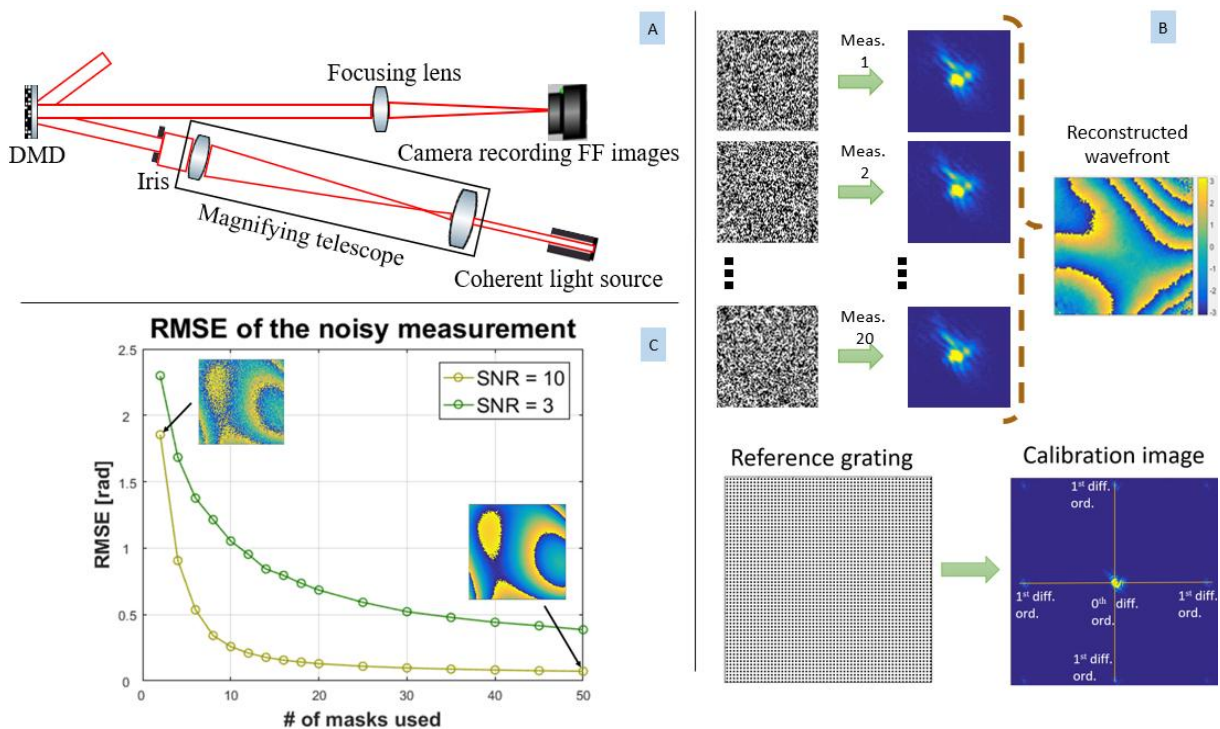

Fig. 1. Wavefront sensing concept and calibration. A) Schematic layout of the experimental setup. The laser beam is expanded and amplitude-modulated by a Digital Micromirror Device (DMD). The modulated beam is focused to form a far-field intensity distribution recorded by a camera. B) Top: Examples of pseudorandom binary amplitude masks displayed on the DMD and corresponding recorded far-field intensity patterns. Bottom: Periodic calibration mask used to determine the area of interest (AOI) on the camera and a representative calibration image. C) Numerically evaluated mean reconstruction root-mean-square error (RMSE) as a function of the number of modulation masks for two representative detector noise levels. The study illustrates the trade-off between measurement redundancy and noise robustness.

The experimental layout of the proposed wavefront sensor is shown schematically in Fig. 1A. In a laser system, a fraction of the laser beam is directed into the diagnostic branch and expanded to uniformly illuminate the active area of the Digital Micromirror Device (DMD). The DMD (DLP® LightCrafter™ 6500 Evaluation Module) is used exclusively as a binary amplitude modulator, operating in an on/off configuration of individual micromirrors. The native resolution of the DMD is 1080x1920 mirrors, each individually addressable by HDMI data output from the GPU. To reduce computational complexity while preserving sufficient spatial sampling for wavefront reconstruction, the DMD surface was grouped into a modulation grid of 120 × 120 super-pixels, where each super-pixel consists of 9 × 9 micromirrors. This choice represents a trade-off between achievable spatial resolution and reconstruction time and is not a fundamental limitation of the method. Two laser sources were used to demonstrate wavelength versatility. For visible measurements, a laser diode operating at $\lambda = 650$ nm was employed. For near-infrared measurements, a Ho-doped fiber oscillator emitting at $\lambda = 2116$ nm was used.

Following amplitude modulation by the DMD, the beam is focused by a lens with focal length $f = 250$ mm, forming a far-field (Fraunhofer) intensity distribution in the focal plane. The intensity distribution is recorded by a camera placed in the focal plane and aligned perpendicular to the optical axis. For visible and near-infrared measurements, different detectors were used to match wavelength sensitivity and signal-to-noise requirements. A CMOS camera (Allied Vision Manta G-125, native resolution 1292 × 964) was used for measurements at 650 nm, while a mid-infrared camera (Tachyon 16k, native resolution 128 × 128) was used for measurements at 2116 nm. Both cameras were operated via Ethernet interfaces, providing sufficient bandwidth for multi-frame acquisition.

Since only the amplitude is modulated, a large fraction of the optical power remains concentrated in the center of the focal spot (see Fig. 1B)) in contrast to phase-modulation-based approaches [50]. For the remainder of this manuscript, we evaluate the acquisition noise as a ratio between the power of the data carrying signal and the power of the noise across the whole area of interest.

To assess the sensitivity of the reconstruction algorithm to detector noise, a numerical study was performed in which synthetic wavefronts were propagated through the forward model and provided with additive noise. The root-mean-square error (RMSE) of the reconstructed phase was evaluated as a function of the number of masks for two levels of signal-to-noise ratio (SNR) of the simulated detector. Figure 1C shows the mean reconstruction error as a function of mask count for two representative noise levels. The RMSE is computed as the standard deviation of the difference between the injected phase and the reconstructed phase after unwrapping.

The experimentally measured background noise levels were approximately 13–18 dB for the visible camera and 5–8 dB for the near-infrared camera, depending on the exposure time. The noise levels used in the simulation intentionally overestimate these values to provide a conservative margin for varying illumination and alignment conditions.

To establish the area of interest (AOI) on the camera corresponding to the spatial bandwidth of the modulated beam, a calibration procedure based on a periodic reference mask was employed. A two-dimensional grid pattern with spatial frequency equal to half the DMD super-pixel frequency was projected onto the DMD (Fig. 1B, bottom). The resulting diffraction pattern in the focal plane contains first-order diffraction maxima whose separation defines the spatial frequency support of the modulated field. The region between these maxima contains the complete spectral information required for reconstruction and is therefore selected as the AOI. This process is easily adapted to any specific configuration of focusing lens and recording camera. The first diffraction maxima along each axis are given by:

$$x_i = \lambda f \nu_i, \tag{6}$$

where $\lambda$ is optical wavelength, $f$ is the focal length of the focusing lens and $\nu_i$ is the spatial frequency of the projected calibration grating along the corresponding axis. The AOI is resampled to match the resolution of the DMD modulation grid so that each recorded far-field image corresponds directly to the power spectrum of the DMD-plane field. Accurate camera alignment and sufficient detector sampling are essential to avoid aliasing of high spatial frequencies. Residual geometric distortions can alternatively be corrected through digital processing prior to the wavefront reconstruction.

Once the AOI is calibrated the beam is sequentially modulated by a set of complementary pseudorandom masks with binary character. Each mask is generated with equal probability of transmitting or blocking individual super-pixels, ensuring approximately uniform average illumination while providing sufficient diversity for phase retrieval. Although recent literature shows that structured and optimized binary masks—such as Hadamard complementary patterns—can reduce noise amplification and lower the number of required measurements, pseudorandom masks were selected in this study to prioritize experimental robustness and simplicity. Preliminary simulations performed under the specific constraints of the present setup indicated increased sensitivity to systematic artifacts for certain structured masks; however, no claim of general superiority of random masks is made.

For the visible experiment, a set of 20 masks was used, while a set of 30 masks was used for near-infrared measurements to compensate the lower detector signal-to-noise ratio. The number of measurements was selected based on a numerical noise analysis described above and is intentionally higher than the theoretical minimum to improve stability under experimental conditions. Wavefront reconstruction was performed using the Reweighted Amplitude Flow algorithm with Optimal Spectral Initialization (RAF-OSI). The algorithm was implemented in MATLAB and operated on the set of recorded far-field intensity images and corresponding binary masks. The reconstructed phase is initially obtained modulo $2\pi$. A phase unwrapping algorithm based on solving the Poisson equation using a discrete cosine transform was applied to recover a continuous wavefront. The reconstruction time for a dataset of 20 masks was approximately 3–5 s on a standard workstation, resulting in an effective sampling rate of approximately 0.2 Hz. The computational cost is dominated by the iterative RAF-OSI updates and was not optimized for real-time operation. The method relies on the sequential acquisition of multiple intensity profiles (20 in the visible experiment), which further limits the achievable temporal bandwidth. Nevertheless, acceleration of the reconstruction could be achieved through a dedicated implementation in a compiled language, exploiting parallelization on GPU, or employing hardware acceleration using FPGA-based processing. As a result, the method in its current status is best suited for quasi-static or slowly varying wavefronts, where high spatial resolution and wavelength flexibility are prioritized over acquisition speed.

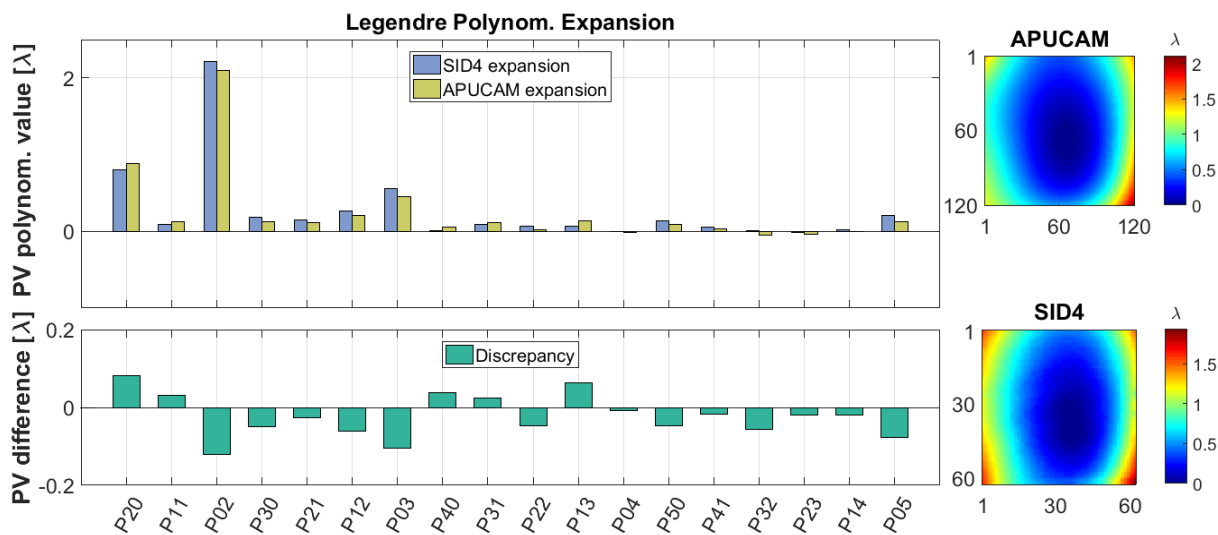

Fig. 2. Comparison of visible-wavelength wavefront reconstruction at λ = 650 nm. Top: Legendre polynomial expansion coefficients (up to 5th order) of the wavefront reconstructed using APUCAM (120 × 120 sampling) and measured using a commercial lateral shearing interferometer (SID4, 62 × 62 sampling). Bottom: Coefficient-wise discrepancy between the two measurements. Piston and tilt terms are omitted due to their dependence on relative alignment. Right: Unwrapped wavefront maps obtained by the two methods. The comparison focuses on consistency of low- and mid-order aberrations rather than absolute metrological accuracy.

As a proof-of-principle, the wavefront reconstructed by the proposed APUCAM method was compared with a measurement obtained using a commercial lateral shearing interferometer (SID4, Phasics S.A.) at 650 nm wavelength. The comparison focuses on assessing consistency in low- and mid-spatial-frequency aberrations rather than establishing absolute metrological equivalence. Figure 2 (right) shows the unwrapped wavefronts measured by APUCAM and SID4, respectively. For quantitative comparison, both wavefronts were expanded into Legendre polynomials up to the fifth order. Piston and tilt terms were excluded, as they depend on the relative alignment and focal spot position and are not intrinsic to the reconstruction fidelity.

The corresponding Legendre coefficients are shown in Fig. 2 (top), while Fig. 2 (bottom) depicts the coefficient-wise discrepancy between the two measurements. Despite the presence of high-spatial-frequency noise in the APUCAM reconstruction, the low-order aberration content shows good agreement between the two methods. This indicates that the proposed approach reliably captures the dominant aberration modes relevant for adaptive optics correction. The comparison presented in Fig. 2 serves to verify that the reconstructed wavefront is physically consistent and suitable for use as a feedback signal in closed-loop correction.

To demonstrate wavelength versatility, the laser source and detector were replaced to perform wavefront reconstruction at λ = 2116 nm. At this wavelength, no commercial wavefront sensor was available for direct comparison. Consequently, validation focuses on internal consistency and physical plausibility of the reconstruction. A binary aperture mask in the form of a simple black-and-white image was superimposed onto the DMD modulation patterns to spatially restrict the region of interest. After AOI calibration and alignment, the wavefront was reconstructed with a resolution of 112 × 112 pixels. Figures 3A and 3B show the reconstructed phase and intensity distributions, respectively. Figure 3C presents an unwrapped phase cut-out corresponding to the imposed aperture, while Fig. 3D illustrates the reconstructed phase in regions where no optical signal is present.

The latter exhibits noise-dominated behavior, as expected in the absence of intensity support. The continuity of the reconstructed phase within the bounded aperture and the absence of spurious structure outside the illuminated region indicate that the algorithm correctly identifies the spatial support of the field. These observations support the feasibility of applying the method in wavelength regimes where conventional wavefront sensors are impractical or unavailable. The primary limitation of the near-infrared reconstruction arises from reduced SNR, inherent to the Mid-IR detector, which amplifies high-spatial-frequency noise in the recovered phase.

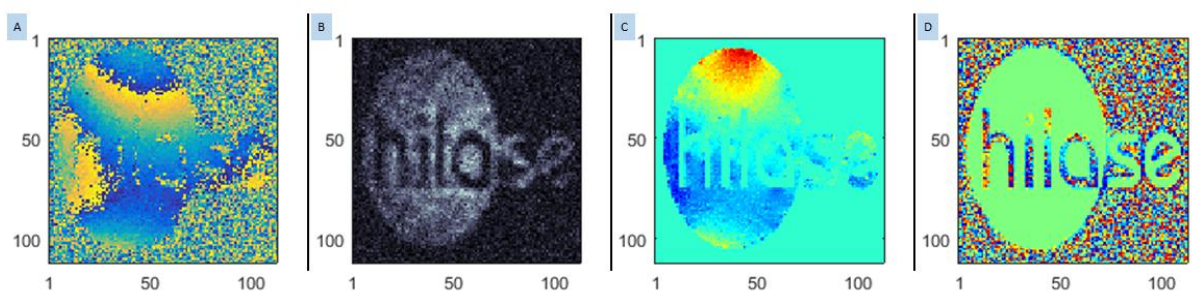

Fig. 3. Near-infrared wavefront reconstruction at $\lambda$ = 2116 nm. A) Reconstructed phase distribution (resolution 112 × 112) within a spatially bounded aperture defined by a superimposed binary mask on the DMD. B) Corresponding reconstructed intensity distribution. C) unwrapped phase cut-out corresponding to the illuminated aperture region. D) Reconstructed phase in regions without optical signal, demonstrating noise-dominated behavior outside the spatial support.

The results illustrate internal consistency of the reconstruction in a wavelength regime where commercial wavefront sensors are limited.

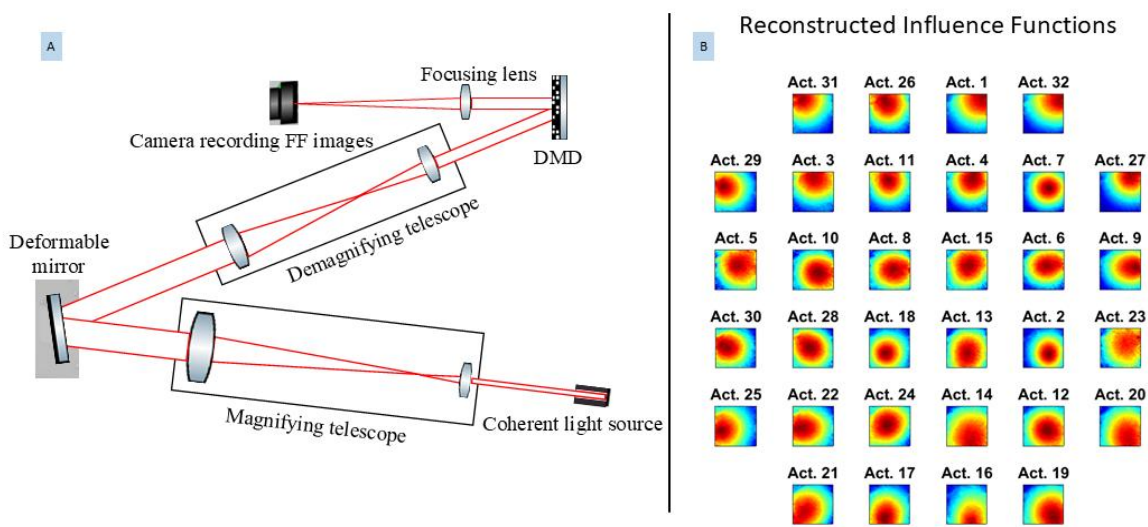

Fig. 4. Closed loop adaptive optics configuration and deformable mirror characterization. A) Simplified schematics of the closed loop setup. The deformable mirror (DM) is imaged onto the DMD plane via a Keplerian relay telescope, ensuring conjugation between correction and sensing planes. B) Experimentally measured influence functions of individual DM actuators reconstructed using APUCAM, illustrating the spatial sampling density and actuator layout.

The intended application of the proposed wavefront sensor is to provide feedback for closed-loop adaptive optics. To assess this capability, a deformable mirror (DM) was incorporated into the visible-wavelength experimental setup ($\lambda$ = 650 nm). A bimorph deformable mirror was placed upstream of the DMD. A Keplerian relay telescope was used to demagnify the beam and image the DM plane onto the DMD plane, ensuring conjugation between the correction and sensing planes. The simplified layout of the closed-loop system is shown in Fig. 4A. The influence functions (IFs) of the deformable mirror actuators were measured using the APUCAM wavefront sensor. The mirror was initially driven with zero bias voltage, and the resulting wavefront was used as a reference. Subsequently, a linear voltage stroke was applied sequentially to each actuator, and the corresponding wavefront response was reconstructed.

A total of 32 influence functions were measured. The resulting response matrix was analyzed using singular value decomposition (SVD), yielding a modal representation of the mirror's controllable aberration space. The spatial structure of the measured influence functions is shown in Fig. 4B. Based on the actuator density and layout, the mirror is capable of reproducing aberrations approximately up to the fifth order of the Legendre polynomial basis. Higher-order aberrations cannot be efficiently corrected due to the discrete spatial sampling of the actuators. Closed-loop wavefront correction was performed by decomposing the residual wavefront reconstructed by APUCAM into the measured influence functions and computing the corresponding actuator voltage updates using a least-squares approach. The voltage increment applied to the i-th actuator is given by

$$V_{n+1}^{(i)} = V_n^{(i)} - g\, V_{LS}^{(i)}, \quad i \in \langle 1, \ldots, n \rangle. \tag{7}$$

Here $V_n^{(i)}$ denote the voltages applied to the $i$-th actuator at the $n$-th iteration, $V_{LS}^{(i)}$ is the least-squares actuator voltage correction obtained from the pseudoinverse of the measured influence-function matrix and the reconstructed residual wavefront, and $g$ is a scalar loop gain.

The closed loop operation with a gain of $g = 0.3$ was used to correct the aberrations of the beam (Fig. 5A, Top inset). The correction loop was terminated once the standard deviation of the reconstructed wavefront exhibited stagnation, typically after 5–8 iterations. A notable limitation arises when targeting a planar wavefront. As the correction proceeds, the focal spot becomes increasingly concentrated, leading to elevated peak intensities. To avoid detector saturation, exposure time must be reduced, which in turn degrades the signal-to-noise ratio of the diffraction features carrying high spatial frequency information. This effect can destabilize the reconstruction near target. To mitigate this issue, a controlled defocus aberration with a peak-to-valley value of 1 $\lambda$ was introduced as an intermediate objective. After correction of higher-order aberrations, the defocus term was analytically removed by projecting it onto the influence function basis and updating the actuator voltages accordingly. This strategy improves numerical stability and is specific to amplitude-only modulation schemes where power concentration cannot be actively redistributed.

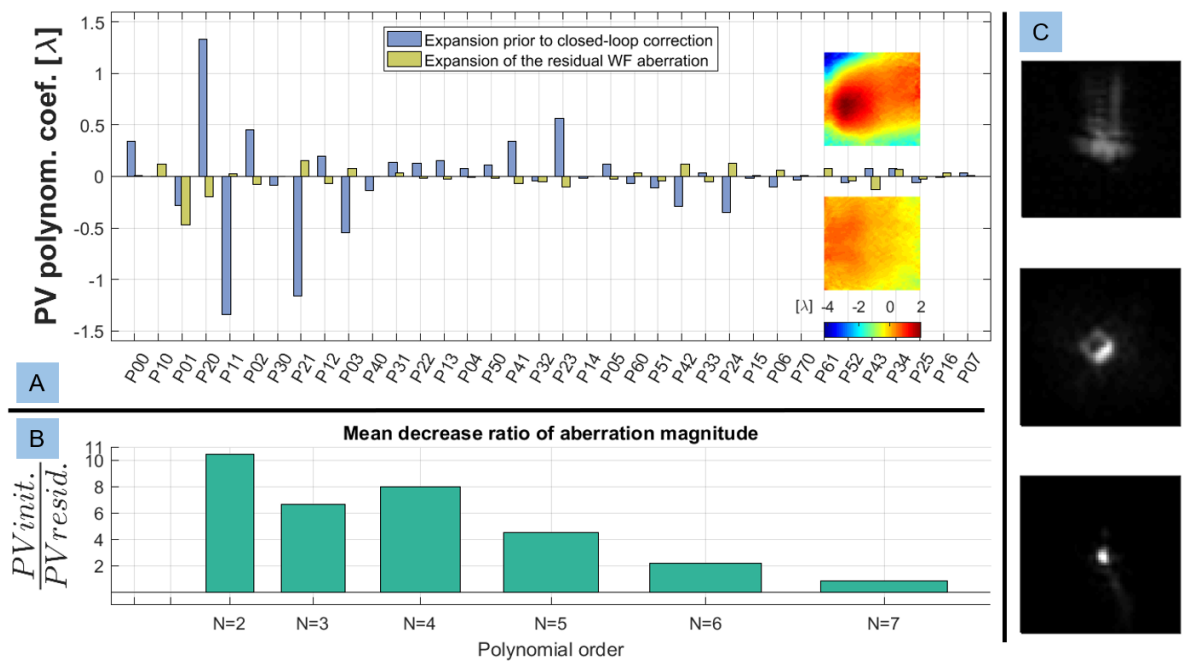

Fig. 5. Closed-loop wavefront correction results at λ = 650 nm. A) Evolution of the reconstructed wavefront during closed-loop optimization. The initial wavefront (RMSE = 1.66 λ, blue Leg. expansion) is progressively corrected over five iterations. A controlled defocus bias is introduced for stability and subsequently removed, yielding a residual wavefront with RMSE = 0.22 λ (yellow Leg. expansion). Insets show representative wavefront maps (resolution 120 × 120). B) Mean improvement ratio of Legendre polynomial coefficients as a function of polynomial order. Aberrations up to 5th order are effectively compensated, while compensation of higher-order modes remains limited due to actuator density. C) Corresponding evolution of the far-field focal spot, showing improvement from a distorted profile (top) to an intermediate defocus aberration (middle) to a corrected, high-Strehl configuration (bottom).

The process of closed loop wavefront correction is shown in Fig 5. Figure 5A illustrates a representative closed-loop correction sequence. The initial wavefront exhibited an RMSE of 1.66 λ. After five correction iterations, the measured residual wavefront RMSE was reduced to 0.22 λ following defocus removal. It should be noted that the reported residual wavefront RMSE of 0.22 λ does not directly represent the physical correction quality of the optical system, as the reconstructed phase exhibits elevated high-spatial-frequency noise, primarily originating from detector noise. These noise components disproportionately contribute to the global RMSE metric, despite having no impact on the low- and mid-order aberrations that dominate beam quality and Strehl ratio. As shown in Fig. 5B, substantial reduction was achieved for aberration orders up to N = 5, while higher-order terms remained largely uncorrected due to actuator density limitations. The evolution of the far-field focal spot during correction is shown in Fig. 5C. The initial distorted focal spot exhibited a Strehl ratio of approximately 0.06. After closed-loop optimization, a Strehl ratio of 0.82 was achieved, corresponding to a wavefront with RMSE of 0.07 λ [51]. This result satisfies the conventional Maréchal criterion [52] SR > 0.8 for a diffraction-limited beam. The sensitivity of RAF-OSI to the noise-induced high-spatial-frequency artifacts reduce loop stability, therefore the ultimate performance is constrained primarily by the emergence of noise during reconstruction and detector dynamic range rather than by the deformable mirror.

## 4. Conclusions

We have presented a reference-free wavefront sensing approach capable of high-resolution reconstruction of laser beam wavefronts, designed to serve as a feedback signal provider for deformable mirrors and related wavefront correction elements. The method is based on binary amplitude modulation using a Digital Micromirror Device and computational phase retrieval, enabling reconstruction of arbitrary wavefront shapes without reliance on wavelength-specific refractive or interferometric components. As a result, the achievable spatial resolution is scalable with the modulator resolution, and the method is not intrinsically limited in dynamic range by the sensing principle.

The performance of the reconstruction algorithm was evaluated under varying noise conditions, allowing estimation of the minimum dataset size required for stable and reliable reconstruction of the complex laser field. In a proof-of-concept experiment, a static wavefront of a $\lambda = 650$ nm laser diode was measured using the proposed APUCAM approach with a spatial resolution of 120 × 120 pixels and compared against a commercial lateral shearing interferometer operating at a resolution of 62 × 62 pixels. Good agreement was observed for low- and mid-order aberrations relevant to adaptive optics correction. The wavelength versatility of the method was further demonstrated by reconstructing a spatially bounded laser beam at $\lambda = 2116$ nm with a resolution of 112 × 112 pixels, a wavelength regime where commercial wavefront sensors are not widely available.

Finally, the proposed wavefront sensor was integrated into a closed-loop adaptive optics experiment incorporating a deformable mirror. The mirror's actuator influence functions were characterized using the APUCAM method, and an initially aberrated wavefront with an RMSE of 1.66 $\lambda$ was successfully corrected to a near-planar wavefront with a residual RMSE of 0.07 $\lambda$. This demonstrates the suitability of the approach for closed-loop wavefront correction in laboratory environments.

The presented method provides an alternative to conventional wavefront sensing techniques, offering scalable spatial resolution, broad wavelength applicability and straightforward experimental implementation. While the computational reconstruction is inherently time-consuming and therefore not suited for rapidly varying wavefronts, the flexibility, simplicity and wavelength agnosticism of the approach make it a valuable diagnostic and adaptive optics tool in applications where measurement speed is secondary to spatial resolution and dynamic range.

**Author Contributions:** Conceptualization, Ondřej Denk, Jan Pilař and Martin Divoký; methodology, Ondřej Denk and Jan Pilař; software, Ondřej Denk; validation, Ondřej Denk and Jan Pilař; formal analysis, Ondřej Denk, Jan Pilař, Martin Divoký and Miroslav Čech; investigation, Ondřej Denk and Jan Pilař; resources, Martin Divoký and Tomáš Mocek; data curation, Ondřej Denk and Jan Pilař; writing—original draft preparation, Ondřej Denk; writing—review and editing, Jan Pilař, Martin Divoký and Miroslav Čech; visualization, Ondřej Denk; supervision, Jan Pilař, Martin Divoký and Miroslav Čech; project administration, Tomáš Mocek; funding acquisition, Martin Divoký and Tomáš Mocek. All authors have read and agreed to the published version of the manuscript.

**Funding:** This article was co-financed by the project LasApp (CZ.02.01.01/00/22_008/0004573) and by the European Union's Horizon 2020 research and innovation program under grant agreement No. 739573. This work was also supported by the Ministry of Education, Youth and Sports of the Czech Republic (Programs NPU I Project No. LO1602, and Large Research Infrastructure Project No. LM2015086).

**Institutional Review Board Statement:** Not applicable.

**Data Availability Statement:** Dataset available on request from the authors.

https://doi.org/10.3390/xxxxx

**Acknowledgments:** During the preparation of this manuscript, the authors used ChatGPT for the purposes of text proofreading. The authors have reviewed and edited the output and take full responsibility for the content of this publication.

**Conflicts of Interest:** The authors declare no conflicts of interest.

## Abbreviations

The following abbreviations are used in this manuscript:

| | |
|---|---|
| MDPI | Multidisciplinary Digital Publishing Institute |
| DOAJ | Directory of open access journals |
| DMD | Digital Micro-mirror Device |
| (N)IR | (Near-)Infrared |
| ER | Error Reduction algorithm |
| APUCAM | Amplitude flow algorithm for Phase retrieval computations Utilizing Coded Aperture Masking |
| RAF-OSI | Reweighted Amplitude Flow with Optimal Spectral Initialization |
| AOI | Area of interest |
| RMSE | Root-mean-square error |
| SNR | Signal-to-noise ratio |
| DM | Deformable mirror |
| IF | Influence function |
| SR | Strehl ratio |